\documentclass{optica-article}

\journal{opticajournal} 

\articletype{Research Article}

\usepackage{lineno}

\begin{document}

\title{Digital-twin imaging based on descattering Gaussian splatting}

\author{Suguru Shimomura,\authormark{*} Kazuki Yamanouchi, and Jun Tanida}

\address{Graduate School of Information Science and Technology, Osaka University, 1-5 Yamadaoka, Suita, Osaka, Japan}

\email{\authormark{*}s-shimomura@ist.osaka-u.ac.jp} 


\begin{abstract*} 
Three-dimensional imaging through scattering media is important in medical science and astronomy. 
We propose a digital-twin imaging method based on Gaussian splatting to observe an object behind a scattering medium.
A digital twin model built through data assimilation, emulates the behavior of objects and environmental changes in a virtual space.
By constructing a digital twin using point clouds composed of Gaussians and simulating the scattering process through the convolution of a point spread function, three-dimensional objects behind a scattering medium can be reproduced as a digital twin.
In this study, a high-contrast digital twin reproducing a three-dimensional object was successfully constructed from degraded images, assuming that data were acquired from wavefronts disturbed by a scattering medium. This technique reproduces objects by integrating data processing with image measurements.
\end{abstract*}

\section{Introduction}
Optical imaging is a fundamental technology for observing objects and is widely applied in various fields including medical science and astronomy.
High-resolution or high-contrast imaging enables the observation of deep tissue conditions {\it in vivo}, providing an understanding of biological mechanisms.
In astronomy, observing distant objects furthers our understanding of the universe. 
However, scattering media distort optical wavefronts,  making it difficult to observe the target objects accurately. 
Obtaining information about objects from optical signals scattered by the environment remains challenging, and several data processing-based methods have been proposed.

Computational imaging is a promising approach for extracting object information from scattered signals by combining optical and computational processing \cite{Mosk2012,Yoon2020,Gigan2022}. 
For example, Richardson--Lucy and Wiener deconvolutions are well-known methods for removing scattering effects from degraded images \cite{Levin2009}. 
Modeling a scattering medium allows for the reproduction of light propagation and the recovery of object structures \cite{Kang2023,Thendiyammal2020}.
Digital holography techniques reconstruct object images from the interference patterns of the speckles and reference beams \cite{Singh2014,Omkuwar2017,Bertolotti2012}.
In addition, speckle correlation and phase retrieval algorithms offer noninvasive imaging for observing objects behind a scattering medium \cite{Katz2014,Okamoto2019,Wang2018,Yeminy2021}.
Another approach involves estimating the transmission matrix that characterizes light propagation \cite{Popoff2011}. 
The transmission matrix can be estimated by measuring the optical responses of the object, and the object can be reconstructed using the matrix inverse function \cite{Popoff2010,Boniface2020}.
Machine learning enables the modeling of scattering processes and reconstruction of object images from speckle patterns \cite{Horisaki2016,Nishizaki2019,Liu2024}. Computational imaging can provide high-contrast object images from scattered signals by employing various retrieval algorithms.
In these methods, it is important to obtain the light-scattering response from a limited spatial region in advance.
The contrast and spatial resolution of the reconstructed image depend on the amount of captured data, and long measurement times are required to extend the field of view behind the scattering medium.
In conventional methods, object reconstruction is performed under the assumption that the scattering medium remains time invariant. 
However, biological and atmospheric turbulence changes dynamically, making it difficult to reconstruct images based on previously obtained data.
In addition, reconstructing three-dimensional (3D) objects requires multiple measurements depending on the cross-sectional view of the reconstructed object. 
Consequently, efficient reconstruction of 3D objects from images degraded by scattering remains a significant challenge.
These issues can be addressed by simultaneously optimizing the measurement and reconstruction by integrating optical and computational processing.

A digital twin constructed through data assimilation is a promising technology for
integrating data acquisition and processing \cite{Tao2022}.
Through simulation using digital-twin dynamics, the environmental and object changes can be emulated in cyberspace. 
The emulation results are analyzed and fed back into the data acquisition method in the physical space.
The digital twin is updated by using the acquired data and utilized for the emulation again. By iterative update, the emulation accurately reproduces the changes in the physical system.
The use of a digital twin not only reproduces, but also predicts object behavior through iterative emulation and analysis, facilitating optimization and decision-making for efficient data acquisition.
By integrating a digital twin into computational imaging, complex light scattering can be modeled, analyzed, and used to efficiently reproduce objects behind the scattering media.

This paper presents a digital-twin imaging method for reproducing 3D objects behind the scattering media.
In the proposed method, a digital twin representing the object behind the scattering medium is constructed using captured images that are degraded by scattering.
By emulating the scattering process in cyberspace and iteratively updating the digital twin, the 3D structure of the object can be reproduced.
Construction of a digital twin requires modeling the target object.
We employed a method that models an object as a collection of points, allowing for flexible adaptation without a predefined spatial resolution.
The optical response of an object can be described as the sum of the signals at each point. Thus, an image of the target object can be obtained if the individual optical responses are accurately reproduced in cyberspace.
To represent the digital twin as point clouds, we employed Gaussian splatting (GS), which represents a 3D space using Gaussians \cite{Kerbl2023}.
In GS, 3D objects are flexibly represented by optimizing the parameters of each Gaussian based on images captured from multiple viewpoints.
Previous studies demonstrated 3D imaging of the human body based on coherent tomography using GS \cite{Li2023,Cai2024}.
Furthermore, combining neural fields with GS enables the creation of realistic 3D spaces without motion blurring \cite{Lee2024,Seiskari2024}.
In the proposed method, the digital twin is constructed and iteratively updated based on a GS algorithm that incorporates the scattering process.
To validate the proposed method, we evaluated the structure of a digital twin constructed from blurred images captured from multiple viewpoints.

\section{Digital twin constructed and updated by descattering Gaussian splatting}
\begin{figure}[tb]
    \centering
    \includegraphics[width=\linewidth]{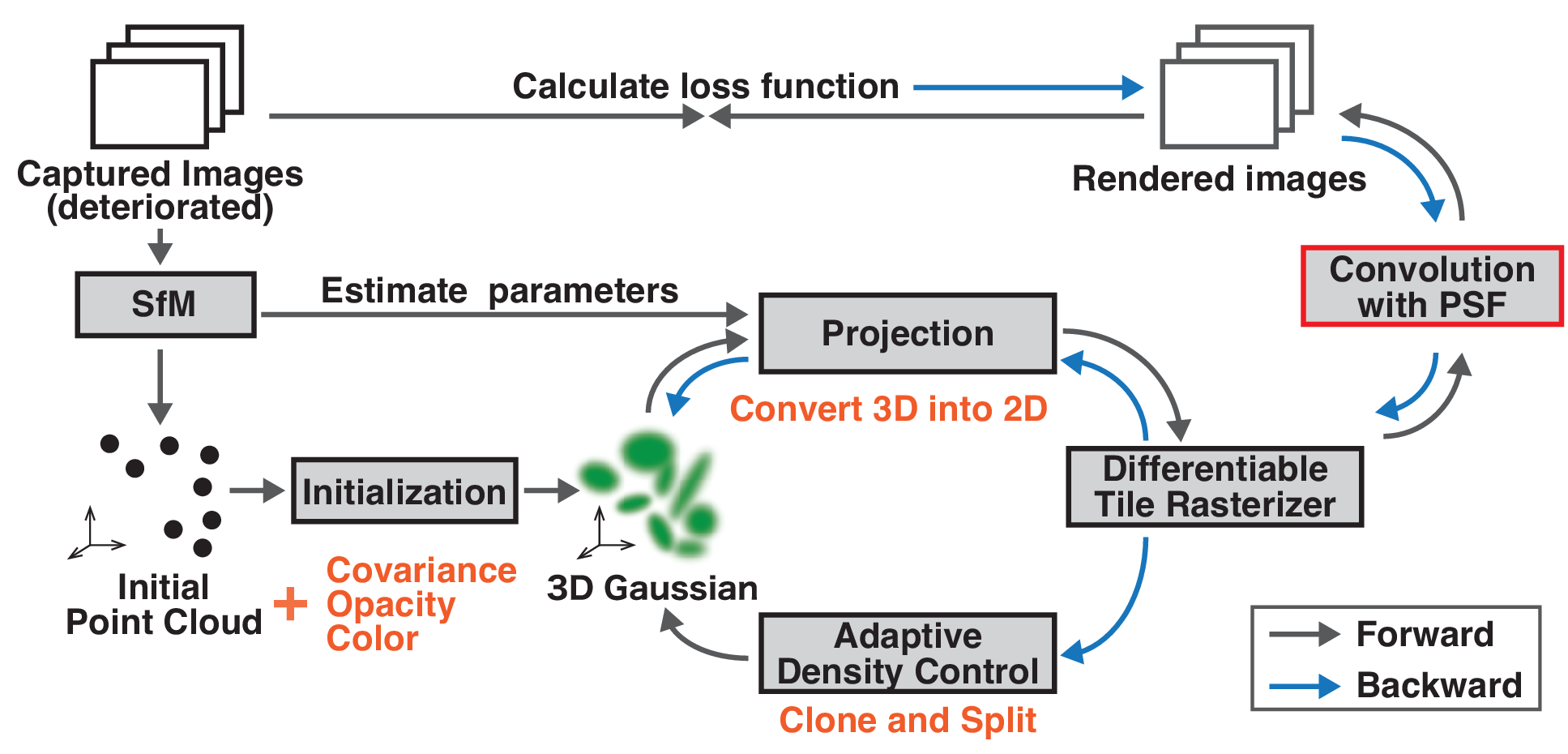}
    \caption{DGS to construct a digital twin of the target object using images degraded by a scattering medium.}
    \label{fig:DGS}
\end{figure}

In GS, 3D space, including various objects, is represented by fitting Gaussians such that their rendered images, observed from multiple viewpoints in cyberspace, correspond to the actual captured images.
To reproduce an object behind a scattering medium from the degraded images, each Gaussian’s parameters are optimized based on the rendered images degraded by a virtually simulated scattering process.
Hereafter,  we refer to GS incorporating the scattering process as descattering Gaussian splatting (DGS).
Figure \ref{fig:DGS} shows the DGS process. 
Initially, the target object behind the scattering medium is captured from multiple viewpoints.
Using structure from motion (SfM), which estimates camera parameters, including position and orientation, point clouds, which are referred to as the digital twin, are generated \cite{Schonberger2016,Snavely2006}.
These points are then converted into Gaussians using the following equation:
\begin{equation}
G(x)= e^{-\frac{1}{2}(x-\mu_i)^T\Sigma_i(x-\mu_i)},
\end{equation}
where $\boldsymbol{x}$ are the spatial coordinates, and $\mu_i$ is the center position of the $i$-th point. 
$\Sigma$ is the 3D covariance matrix that represents the shape of the Gaussian.
Individual Gaussians have parameters $\alpha$ and spherical harmonic coefficient $c$, which represent their opacity and color, respectively. 
The Gaussians are then projected onto a 2D image by rendering.
During the projection, the 3D covariance matrix  $\Sigma$ \cite{zwicker2001} is converted into a 2D covariance for rendering.
After converting the parameters of the individual Gaussians to 2D, the images are rendered by a differentiable tile rasterizer using the positions and poses estimated by SfM.
Each pixel value $I_{\rm pix}$ in the rendered image $I_{GS}$ is formulated as an alpha blending of $N$ ordered points that overlap the pixel:
\begin{equation}
I_{\rm pix} = \Sigma_i^N c_i\alpha_i^{2D}\prod_j^{i-1}(1-\alpha_j^{2D}),
\end{equation}
where $\alpha_i^{2D}$ is the opacity of the $i$-th Gaussian weighted by the 2D Gaussian covariance $\Sigma^{2D}$ to $\alpha_i$. 
To construct a digital twin for modeling the target object behind a scattering medium using Gaussians, the scattering and optimization processes must be integrated.

In the proposed method, the rendered images $I_{GS}$ are virtually degraded into $I_{deter}$ by convolving them with a point spread function (PSF) representing light propagation:
\begin{equation}
    I_{deter} = {\rm PSF}\otimes I_{GS},
\end{equation}
where $\otimes$ denotes a convolution operation. 
The use of PSF allows the simulation of the scattering process in cyberspace, thereby enabling the construction of a digital twin of a 3D object behind the scattering medium.
The loss function $L$ was computed using the mean absolute error (MAE) and structural similarity index measure (SSIM) between the rendered images after convolution and the captured images.
Each Gaussian’s parameters are updated through backward processing of the loss function via deconvolution of virtually degraded images. In adaptive density control, Gaussians are split, cloned, or removed based on the gradient of the loss function $L$.
Through optimization process, the digital twin, which consists of Gaussians, is continuously updated using both virtual and actual images. Finally, the emulation results in cyberspace represent the actual them in real space, and therefore, the digital twin can reproduce the object behind the scattering medium.

\section{3D object reproduction by digital-twin imaging based on DGS}
To evaluate the performance of the proposed method, we reproduced a 3D object as a digital twin from blurred images. 

\subsection{Experimental setup}
\begin{figure}
    \centering
    \includegraphics[width=10cm]{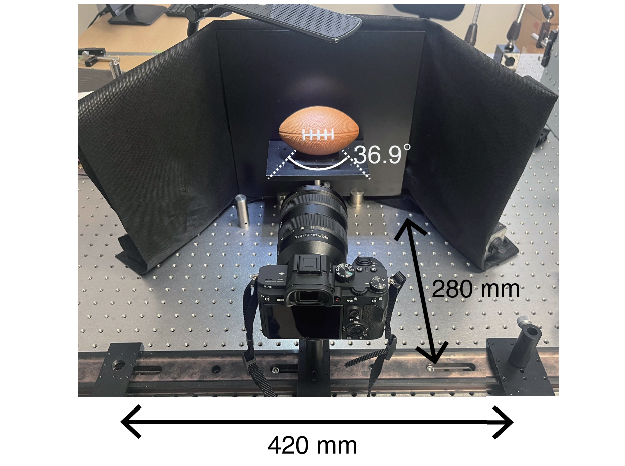}
    \caption{Experimental setup for imaging the target object from various viewpoints.}
    \label{fig:expsetup}
\end{figure}

Figure \ref{fig:expsetup} shows the experimental setup.
The target object was placed at the center of a table, and a rail was positioned along its side.
The target object was imaged using a CMOS camera (Sony $\alpha$7I\hspace{-1.2pt}I\hspace{-1.2pt}I). The camera moved along the rail, covering a length of 420 mm, corresponding to a 36.6 $\circ$  field of view.
To validate the proposed method, 
images degraded by the scattering medium were virtually generated by convolving the captured images using a PSF.
Assuming a scattering process, the PSF was modeled as a 2D Gaussian distribution with a standard deviation of five pixels.

\subsection{Experimental results using digital twin constructed by DGS}
\begin{figure}
\includegraphics[width=\linewidth]{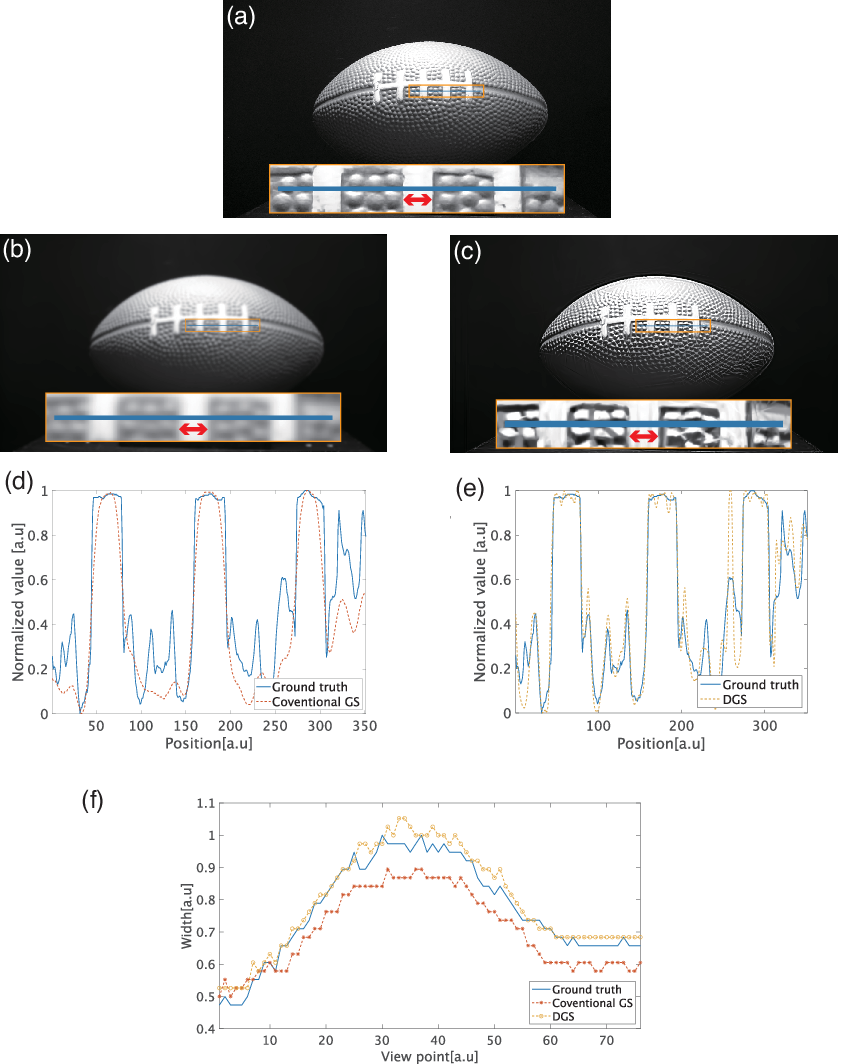}
\caption{
(a) An image captured from a single viewpoint. Rendered images of the digital twins using (b) conventional GS and (c) proposed DGS. The bottom sections of (b) and (c) show enlarged views of the areas highlighted by orange squares.
(d, e) Pixel value profiles along the blue lines shown in (b) and (c), respectively. (f) Linewidths along the red lines with arrows in (b) and (c) from various viewpoints.}
\label{fig2}
\end{figure}

A rugby ball with an uneven surface was used as the target object [Fig. \ref{fig2}(a)].
To assess the performance of the proposed method, digital twins were constructed from blurred images using both the conventional GS and the proposed DGS.

Figures \ref{fig2}(b) and (c) show the digital twins constructed using the conventional GS and the proposed DGS, respectively.
A total of 76 images were utilized as inputs to SfM within each framework.
Rendered views of each digital twin from various angles are provided in Supplementary Material Movie S1.
The surface of the ball constructed using the conventional GS was blurred [Fig. \ref{fig2}(b)].
By contrast, the topography of the digital twin constructed using DGS was clearly reproduced [Fig. \ref{fig2}(c)].
Figures \ref{fig2}(d) and \ref{fig2}(e) show pixel intensity profiles along the blue lines in Figures \ref{fig2}(b) and \ref{fig2}(c), respectively.
With the conventional GS, the regions representing the uneven surface of the ball were not reproduced, although the white lines were preserved.
The proposed method successfully constructed the digital twin with an uneven surface that closely corresponded to the ground truth, as measured from the actual images of the rugby ball.
These results demonstrate that integrating the scattering and construction processes allows the digital twin to retain fine surface details, even when generated from blurred images.

To evaluate 3D structural reproducibility, the widths of the red lines with arrows were measured from virtually generated images captured from multiple angles.
Figure \ref{fig2}(f) shows the measured linewidths from different viewpoints.
The widths of the digital twin constructed using the proposed method closely matched those obtained from the actual images.
The concordance ratio between the widths of the captured and rendered images was 26.3 for DGS, compared with 3.95 for conventional GS.
These results confirm that the proposed method effectively reproduces 3D objects with high fidelity.

By utilizing a digital twin, the back side of an object, where no direct measurement data exist, can be estimated. This estimation provides feedback for further measurements.
The loss function also facilitates the refinement of actual camera parameters such as position and orientation.
Captured images, informed by feedback from the digital twin, are effectively used to optimize the parameters of the Gaussians representing the object.
The loss function is iteratively applied to refine the camera parameters.
This iterative process of optimizing the measurement conditions enables efficient acquisition of 3D object information while reducing the number of images required.

\section{Conclusion}
In this paper, we propose a digital-twin imaging method to model and reproduce the target object using DGS. The convolution of PSF integrated into GS reproduces the scattering process during image capture.
The experimental results demonstrated that the topography of the digital twin constructed from the blurred images corresponded to that of the real object. Our findings suggest that digital-twin imaging can serve as an effective approach for observing objects behind scattering media.

\begin{backmatter}
\bmsection{Funding} 
JSPS KAKENHI (Grant Number 20H05890, 25K21339).

\bmsection{Disclosures}
The authors declare no conflicts of interest.

\bmsection{Data availability} 
Data underlying the results presented in this paper are not publicly available at this time but may be obtained from the authors upon reasonable request.

\end{backmatter}

\bibliography{MyCollection}

\end{document}